# A transition in the spectral statistics of quantum optical model by different electromagnetic fields

H. Sabri[*]

Department of Physics, University of Tabriz, Tabriz 51664, Iran.

---
[*] E-mail: h-sabri@tabrizu.ac.ir

## Abstract

The effects of the quantized electromagnetic fields on the spectral statistics of two-level atoms are considered. We have used the Berry- Robnik distribution and Maximum Likelihood estimation technique to analyze the effect of mean photon and two level atoms numbers and also the quantum number of considered states. Results suggest the intensity of electromagnetic fields as control parameter which the spectral statistics change due to its variation in the 0-1200 region. Also, we observed the universality in the spectral statistics of considered systems when the number of two level atoms is approached to unrealistic limits and there are some suggestions about the effect of the quantum number of selected levels and the atom-field coupling constant on level statistics.

## Introduction

The microscopic many-body interaction of particles in Fermi systems is rather complicated. Several theoretical approaches to the description of the Hamiltonian which are based on the statistical properties of its discrete levels are applied for solutions of realistic problems. For a quantitative measure for the degree of chaoticity in the many-body forces, the statistical distributions of the spacing between the nearest-neighboring levels were introduced in relation to the so called Random Matrix Theory (RMT). The fluctuation properties of quantum systems with underlying classical chaotic behavior and time reversal symmetry correspond with the predictions of the Gaussian orthogonal ensemble (GOE) of random matrix theory. On the contrary, integrable systems lead to level fluctuations that are well described by the Poisson distribution, i.e., levels behave as if they were uncorrelated [1-15]. The information on regular and chaotic nuclear motion available from experimental data is rather limited, because the analysis of energy levels requires the knowledge of sufficiently large pure sequences, i.e., consecutive levels sample all with the same quantum numbers $(J, \pi)$ in a given nucleus. This means, one needs to combine different level schemes to prepare the sequences and perform a significant statistical study.

An interesting example for quantum chaos field is the most basic quantum optical model of a two level atom which interacting with a single mode of a quantized electromagnetic field. Such systems under some special initial conditions such as strong coupling between the field and atom show irregularity which is inconsistent with the results of Graham et al [16] and Ku [17]. In this paper, we have used the formalism of Rui-hua et al which is introduced in Ref.[18] to determine the energy levels of considered systems. With employing the technique which is introduced in Refs.[9-12], we have studied the effect of different electromagnetic fields on spectral statistic of the systems with emphasis on the nearest neighbor spacing statistics.

## 2. Method of analysis

The spectral fluctuations of nuclear levels have been analyzed by different statistics which are based on the comparison of the statistical properties of nuclear spectra with the predictions of Random Matrix Theory (RMT) [1-2]. The Nearest neighbor spacing distributions (NNSD), or $P(s)$ functions, is the observable most commonly used to study the short-range fluctuation properties in nuclear spectra. NNSD statistics requires complete and pure level scheme which these condition are available for a limited number of nuclei. These requirements force us to combine the level schemes of different atoms to construct sequences. On the other hand, one must unfold the considered sequence which means each set of energy levels must be converted to a set of normalized spacing. To unfold our spectrum, we had to use some levels with same symmetry. This requirement is equivalent with the use of levels with same total quantum number ($J$) and same parity. For a given spectrum $\{E_i\}$, it is necessary to separate it into the fluctuation part and the smoothed average part, whose behavior is nonuniversal and cannot be described by RMT [1]. To do so, we count the number of the levels below $E$ and write it as

$$N(E) = N_{av}(E) + N_{fluct}(E) \quad , \tag{2.1}$$

Then we fix the $N_{ave}(E_i)$ semiclasically by taking a smooth polynomial function of degree 6 to fit the staircase function $N(E)$. The unfolded spectrum is yield with the mapping

$$\{\tilde{E}_i\} = N(E_i) \quad , \tag{2.2}$$

This unfolded level sequence, $\{\tilde{E}_i\}$, is dimensionless and has a constant average spacing of 1 but actual spacing exhibits frequently strong fluctuation. Nearest neighbor level spacing is defined as $s_i = (\tilde{E}_{i+1}) - (\tilde{E}_i)$. Distribution $P(s)$ will be in such a way in which $P(s)ds$ is the probability for the $s_i$ to lie within the infinitesimal interval $[s, s+ds]$. The NNS probability distribution function of nuclear systems which spectral spacing follows the Gaussian Orthogonal Ensemble (GOE) statistics is given by Wigner distribution [1]

$$P(s) = \frac{1}{2}\pi s e^{-\frac{\pi s^2}{4}} \quad , \tag{2.3}$$

This distribution exhibits the chaotic properties of spectra. On the other hand, the NNSD of systems with regular dynamics is generically represented by Poisson distribution

$$P(s) = e^{-s} \quad , \tag{2.4}$$

It is well known that the real and complex systems such as nuclei are usually not fully ergodic and neither are they integrable. Different distribution functions [20-24] have been proposed to compare the spectral statistics of considered systems with regular and chaotic limits quantitatively and also explore the interpolation between these limits [12-15]. Berry- Robnik distribution [13] is one of the popular distributions

$$P(s,q) = [q + \frac{1}{2}\pi(1-q)s] \times \exp(-qs - \frac{1}{4}\pi(1-q)s^2) \quad , \tag{2.5}$$

This distribution is derived by assuming the energy level spectrum is a product of the superposition of independent subspectra which are contributed respectively from localized eigenfunctions into invariant (disjoint) phase space and interpolates between the Poisson and Wigner with $q=1$ and 0, respectively. To consider the spectral statistics of sequences, one must compare the histogram of sequence with Berry-Robnik distribution and extract it's parameter via estimation techniques. To avoid the disadvantages of estimation techniques such as Least square fitting (LSF) technique which has some unusual uncertainties for estimated values and also exhibit more approaches to chaotic dynamics, Maximum Likelihood (ML) technique has been used [10] which yields very exact results with low uncertainties in comparison with other estimation methods. The MLE estimation procedure has been described in detail in Refs.[9-12]. Here, we outline the basic ansatz and summarize the results. In order to estimate the parameter of distribution, Likelihood function is considered as product of all $P(s)$ functions

$$L(q) = \prod_{i=1}^{n} P(s_i) = \prod_{i=1}^{n} [q + \frac{1}{2}\pi(1-q)s_i] e^{-qs_i - \frac{1}{4}\pi(1-q)s_i^2} \quad , \tag{2.6}$$

The desired estimator is obtained by maximizing the likelihood function, Eq.(2.6),

$$f: \sum \frac{1 - \frac{\pi s_i}{2}}{q + \frac{\pi}{2}(1-q)s_i} - \sum (s_i - \frac{\pi s_i^2}{4}) \quad , \tag{2.7}$$

We can estimate "$q$" by high accuracy via solving the above equation by the Newton-Raphson method

$$q_{new} = q_{old} - \frac{F(q_{old})}{F'(q_{old})} \quad .$$

which is terminated to the following result

$$q_{new} = q_{old} - \frac{\sum \frac{1 - \frac{\pi s_i}{2}}{q_{old} + \frac{\pi s_i}{2}(1-q_{old})} + \sum s_i + \frac{\pi s_i^2}{4}}{\sum \frac{-(1 - \frac{\pi s_i}{2})^2}{(q_{old} + \frac{1}{2}\pi(1-q_{old})s_i)^2}} \quad , \tag{2.8}$$

In ML-based technique, we have followed the prescription was explained in Ref.[9], namely maximum likelihood estimated parameters correspond to the converging values of iterations Eq. (2.8) which for the initial values we have chosen the values of parameters were obtained by LS method.

## 3. Two level atom interacting with a single mode of a quantized electromagnetic field

The Hamiltonian and the procedure which must apply to determine the energy levels for two level atom interacting with a single mode of a quantized electromagnetic field has been introduced and described in detail in Refs.[16-18]. Here, we outline the basic ansatz and summarize the results. The Hamiltonian of system which is considered here is [18]

$$H = \Omega a^\dagger a + \omega S_z + 2G(a^\dagger + a)S_x \qquad , \qquad (3.1)$$

where $S_z = \frac{1}{2}\sum_{i=1}^{N_A} \sigma_{z,i}$, $S_x = \frac{1}{2}\sum_{i=1}^{N_A} \sigma_{x,i}$, $\sigma$'s are Pauli spin matrices for individual two level atom. $N_A$ is the total number of two level atoms and $\omega$ is the atomic transition frequency. Also, $a^\dagger$ and $a$ are the creation and annihilation operators of the field with frequency $\Omega$ which the communication relation between them is $[a, a^\dagger] = 1$. $G$ denotes the atom-field coupling constant. In order to study the effect of the mean photon number $\bar{N}_F$ to describes the strength of electromagnetic field and $N_A$, namely the number of atoms, the Hamiltonian (1) has considered as following

$$H = \bar{N}_F \Omega \left(\frac{a^\dagger a}{\bar{N}_F}\right) + N_A \omega \left(\frac{S_z}{N_A}\right) + 2G N_A \sqrt{\bar{N}_F} \left(\frac{S_x}{N_A}\right)\left(\frac{a^\dagger + a}{\sqrt{\bar{N}_F}}\right) \qquad , \qquad (3.2)$$

where $S_i / N_A$, $i = x, y$ and $z$, $(a^\dagger + a)/\sqrt{2\bar{N}_F}$ and $i(a^\dagger - a)/\sqrt{2\bar{N}_F}$ are the dynamical variables satisfying the following communication relations[18],

$$[(\frac{S_i}{N_A}),(\frac{S_j}{N_A})] = i\frac{1}{N_A}\sum_k \varepsilon_{ijk}(\frac{S_k}{N_A}) \qquad , \qquad [\frac{a^\dagger + a}{\sqrt{2\bar{N}_F}}, i\frac{(a^\dagger - a)}{\sqrt{2\bar{N}_F}}] = i\frac{1}{\bar{N}_F} \qquad (3.3)$$

For such systems, the different $N_A$ and $\bar{N}_F$ values when $\bar{N}_F\Omega$, $N_A\omega$ and $2GN_A\sqrt{\bar{N}_F}$ are kept fixed, cause differences due to the changes in the Eqs. (8) and (9). For example, when $1/N_A$ and $1/\sqrt{\bar{N}_F}$ approach zero, equations (8) and (9) turn into classical Poisson brackets which are the same result are expected for classical limit. We have used the same method in the diagonalization of Hamiltonian (3.2) which are introduced and applied in the Refs.[16-18]. This means, we have used the $|m,n\rangle$ as the basis of space which the effect of different operators of considered Hamiltonian on these sates are

$$S_z|m,n\rangle = m|m,n\rangle \qquad\qquad a^\dagger a|m,n\rangle = n|m,n\rangle \qquad (3.4)$$

where $m$ is in the range $\sim N_A/2$ and $n \geq 0$ and integer. A numerical method to truncating an infinite matrix to finite order has applied [16-18] to evaluate the energy eigenvalues.

## 4. Results

In this paper, we look the effect of electromagnetic field's intensity and also the number of two level atoms on spectral statistics of optical model. To this aim, with using different values for parameters of Hamiltonian which are listed in the Table 1, we have determined the lowest 250 levels with the same $m$ values. These energy levels are grouped in different sequences which are labeled with the various $N_A$ and $\bar{N}_F$ values. These sequences are unfolded and then analyzed via maximum likelihood estimation technique in the Berry- Robnik distribution framework. Since, the exploration of the majority of short sequences yields an overestimation about the degree of chaotic dynamics which are measured by distribution parameters, i.e. $q$, therefore, we would not concentrate only on the implicit values of these quantities and examine a comparison between the amounts of this quantity in each sequences.

Similar analyses have been done for some quantum optical models which describe two level atoms. These studies illustrate the statistical situation of different systems in related to chaotic or regular behavior by comparison of the histogram of considered sequence and the curves of these two limits in both NNSD or $\Delta_3(L)$ statistics. These analyses cannot compare the chaocity (or regularity) difference between several sequences and therefore, it is not available to perform a complete investigation for describing the relation between different parameters and statistical properties.

Table1. Parameters of Hamiltonian (3.2) which some of them have borrowed from Ref.[18]. $q$ values describe the chaocity degrees of each systems.

|  | $\Omega/\omega$ | $G/\sqrt{\Omega\omega}$ | $N_A$ | $\bar{N}_F$ | $q$ |
|---|---|---|---|---|---|
| set (I)   | 1    | 0.2  | 21 | 0-40  | 0.52±0.09 |
| set (II)  | 1    | 0.2  | 1  | 0-200 | 0.83±0.12 |
| set (III) | 1    | 0.09 | 21 | 0-40  | 0.51±0.08 |
| set (IV)  | 1    | 1    | 21 | 0-40  | 0.49±0.13 |
| set (V)   | 0.09 | 0.3  | 21 | 0-110 | 0.70±0.10 |
| set (VI)  | 9    | 0.3  | 21 | 0-28  | 0.41±0.08 |

These results show an obvious relation between the mean photon number, $\bar{N}_F$ which describe the intensity of electric field, and the chaocity degree of each system. The most regular dynamic has assigned for system that describe via the parameters of set (II) with the maximum value of $\bar{N}_F$. Also, these results propose same chaocity degree for such systems which have the same $N_A$ and $\bar{N}_F$ values but other parameters of Hamiltonian are changed for them.

We also, considered a special case which the other quantities of Hamiltonian, $\Omega/\omega$; $G/\sqrt{\Omega\omega}$, are assumed to be fix as reported in Table 1 and we have changed the $\bar{N}_F$ and $N_A$ values. As have showed in the Figures 1 and 2, respectively, we have changed the mean photon number in the range 0-1200 and consider the variation of two level atom numbers in the range 0-200 and.

In Figure1, we have observed a deviation of GOE limit (chaocity) and tend to regular dynamics (in all considered systems with different sets of Hamiltonian parameters) when the $\bar{N}_F$ values reach to 700. On the other hand, when we increase the $\bar{N}_F$ values in the range 700-1200, an opposite behavior has happened. In this region, the spectral statistics of considered systems approach to chaotic dynamics similar to what happened in the smallest values of $\bar{N}_F$. One may consider, when the magnetic field is sufficiently strong, around the Dirac point where the density of states is low, the energy levels are quantized into Landau levels and exhibit deviation from Poisson to GOE limits.

Our results in Figure 2 show universality in the statistical properties of considered systems. In different sets of Hamiltonian parameters, when the number of two level atom increase and tend to $N_A$ =200, the regularity, *e.g.* Poisson limit or $q \to 1$, is the dominant behavior in the spectral statistics. The apparent regularity for different systems in this condition, may report that, the identity of atoms make impossible to define the rotation for these systems. These results similar to the predictions of Paar *et al* in [46-47], confirm, the rotation of systems contribute to the suppression of their chaotic dynamics.

On the other hand, the transition in the spectral statistics due to the two level numbers is observed, too for two set of systems, namely (II) and (VI). In transitional region, system goes from a symmetry limit to another dynamical symmetry limit and therefore, a symmetry broken occurs to sub-algebra. It means, combinations of different symmetries visualized by atomic systems and consequently, one can expect a deviation of regular dynamics in transitional regions.

Similar transitions are observed in the quantum phase transitional region between different dynamical symmetry limits of algebraic structures which for example is observed in nuclei [25-29]. For such systems, sudden changes in structure, which are independent of temperature, control by nucleon number as control parameter. Our results for same effect due to photon number or two level numbers suggest similar meaning to these parameters but for an overall summary, more general description on atomic systems are required [25-31].

We also look about the effect of the quantum number of considered states in the spectral statistics. For this investigation, we have fixed other quantities of Hamiltonian as $\Omega/\omega=1$, $G/\sqrt{\Omega\omega}=1$, $N_A= 21$ (a realistic case) and we changed the $\bar{N}_F$ in the 700-1200 region which as previous results, we have observed an approach to chaotic dynamics in all systems. We have determined the eigenvalues of Hamiltonian (3.2) on states with different *m* values. As have displayed in Figure 3, an obvious relation between the chaocity of considered systems and quantum number of considered states can realize where an approach to more chaotic dynamics are report for such systems with bigger *m* values. Also, the rate of this propensity is more for the states with the biggest *m* values. As have mentioned in Ref.[32], the relatively weak strength of spin-orbit interaction is unable to destroy the regular single–particle mean–field motion completely. These mean, one may conclude that, when the *m* values are increased, the strength of interaction yield a more chaotic dynamics. This chaocity also may be related to the strength of pairing force in comparison with spin-orbit interaction force but for a significant conclusion, we need to consider more general cases.

## 5. Summary and conclusion

We look about the spectral statistics of quantum optical model in different conditions. For this aim, Berry- Robnik distribution and also MLE technique have been employed to describe the statistical situation of different systems. The difference in the chaocity parameter of each sequence is statistically significant. The obvious changes in the spectral statistics due to the variation of mean photon and also two level atom numbers are observed and suggest a control parameter like meaning to these quantities. Also, the universality in the approach of spectral statistics to regular dynamic is reported when the number of two level atoms tends to 200.

## Acknowledgement

This work is published as a part of research project supported by the University of Tabriz Research Affairs Office.

## References

[1]. H. A. Weidenmuller, G. E. Mitchel, Rev. Mod. Phys. 81 (2009) 539.
[2]. C.W.J. Beenakker, To appear in The Oxford Handbook of Random Matrix Theory,Oxford University Press 2011, editors G. Akemann, J. Baik, and P. Di Francesco.
[3]. J. M. G. GÓMEZ *et al*. Phys. Rep 499 ( 2011) 103.
[4]. R. Graham, M. Hohenerbach. Z. Phys. B. 57 (1984) 233.
[5]. A. Y. Abul-Magd *et al*. Ann. Phys. (N.Y.) 321 (2006) 560.
[6]. A. Relano, J. M. G. Gomez, R. A. Molina, J. Retamosa, E. Faleiro, Phys. Rev. Lett. 89 (2002) 244102.
[7]. Y. Alhassid. Rev. Mod. Phys. 72 (2000) 895.
[8]. F. M. Izrailev, Phys. Rep. 196 (1990) 299.
[9]. M. A. Jafarizadeh *et.al*. Nucl. Phys. A. 890-891(2012) 29.
[10]. R. Xie, D. Liu, G. Xu, Z. Phys. B. 99 (1996) 605.
[11]. R. Graham, M. Hohenerbach, Phys. Lett. A, 101 (1984) 61.
[12]. Ku, M, Phys. Rev. Lett. 54 (1985) 1343.
[13]. S. Swain. Phys. Lett. A 39 (1972) 143.
[14]. H. Hirooka *et al*, Phys. Lett. A 101 (1984) 115.

[15]. M. Schmutz. Phys. Lett. A 103 (1984) 24.
[16]. T. A. Brody. Lett. Nuovo Cimento, 7 (1973) 482.
[17]. C. H. Lewenkopt *et al*, Phys. Lett. A 155 (1991) 113.
[18]. A. Maciejewski, M. Przybylska, T. Stachowiak, Phys. Lett. A 378 (2014) 16.
[19]. Z. Harsij *et al*, Phys. Rev. A, 86 (2012) 063803.
[20]. P. W. Milonni, J. R. Ackerhalt, H. W. Galbraith, Phys. Rev. Lett, 50 (1983) 966.
[21]. A. Nath, D. S. Ray, Phys. Lett. A, 116 (1986) 104.
[22]. M.V. Berry, M. Robnik. J. Phys. A. 17 (1984) 2413.
[23]. H. Sabri *et al*, Random Matrices: Theory and Applications, 3 (2014) 1450017.
[24]. L. Huang, Y. Lai, C. Grebog. Phys. Rev. E, 81 (2010) 055203.
[25]. P. P´erez-Fern´andez *et al*, Phys. Rev. E, 83 (2011) 046208.
[26]. K. Zyczkowski, G. Lenz, Z. Phys. B. 82 (1991) 299.
[27]. M. Serbyn, J. E. Moore, Phys. Rev. B, 93 (2016) 041424.
[28]. H. Venzl, A. J. Daley, F. Mintert, A. Buchleitner, Phys. Rev. E 79 (2009) 056223.
[29]. D. Fischer, D. Hoffmann, S. Wimberger, Phys. Rev. A 93 (2016) 043620.
[30]. Vinayak, M. Žnidarič, Journal of Physics A: Mathematical and Theoretical, 45 (2012) 125204.
[31]. K. Sacha, J. Zakrzewski, Phys. Rev. Lett, 86 (2001) 2269.
[32]. H. Sabri, Nucl. Phys. A, 941 (2015) 364.

# Figure caption

Figure1. The chaocity degrees, *q* values, for different systems that are described by the sets of Hamiltonian's parameters which are presented in Table 1. For two region, the absence of magnetic field and strong ones, chaotic dynamics is observed for considered systems.

Figure2. Universality in the spectral statistics of different systems is observed when the number of two level atoms approached to 200. Also, the transitional behavior is observed for two sets of considered systems ( set (II) and (VI)) due to the different $N_A$ values.

Figure3. Spectral statistics of different systems are labeled by different *m* values. Results suggest more chaotic dynamics in the same condition for systems have bigger *m* values and also the rate of this approach is more for such systems.

Fig.1.

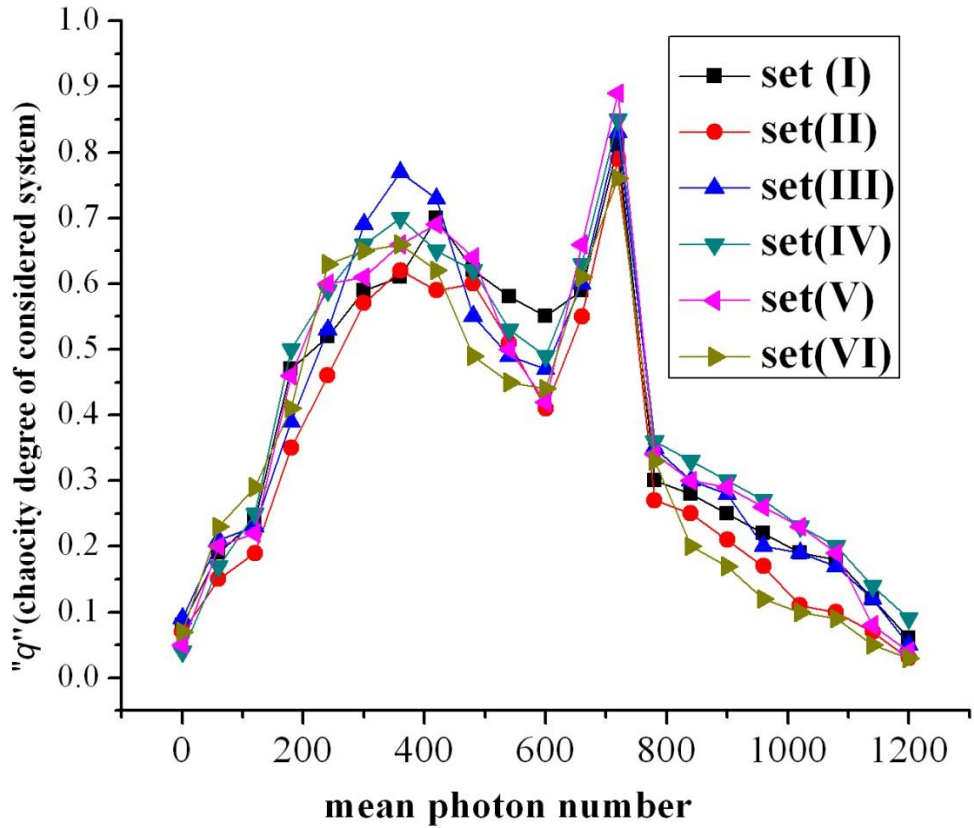

Fig2.

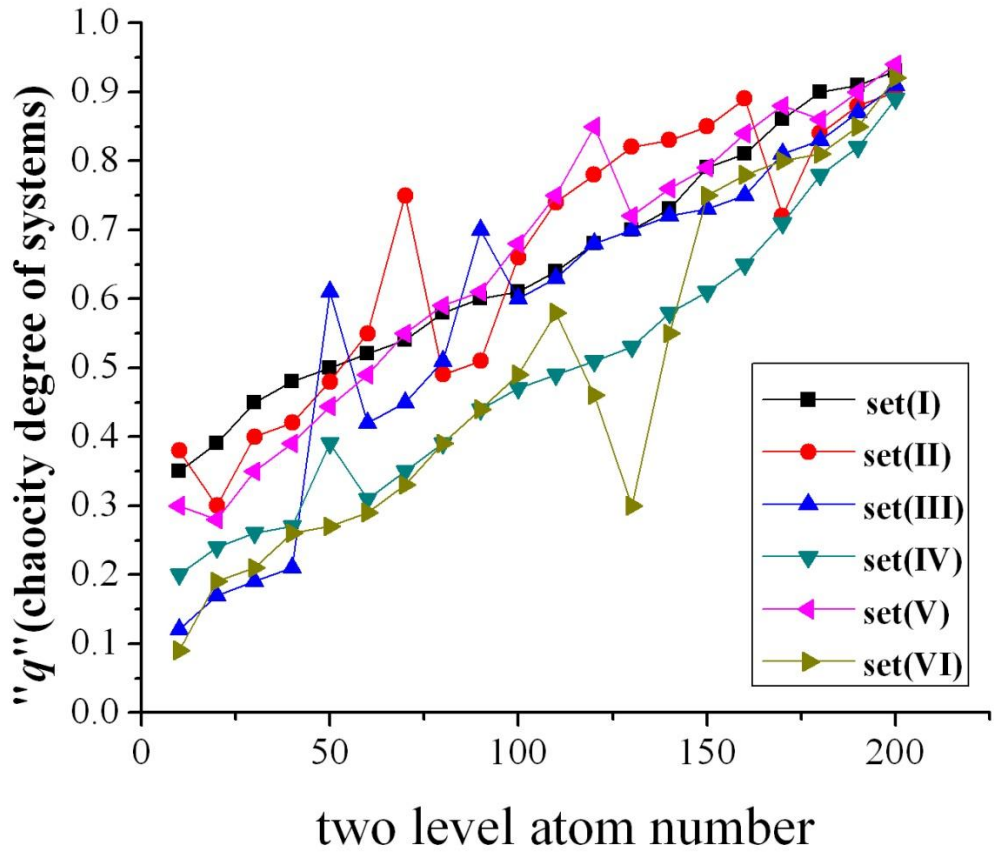

Fig3.

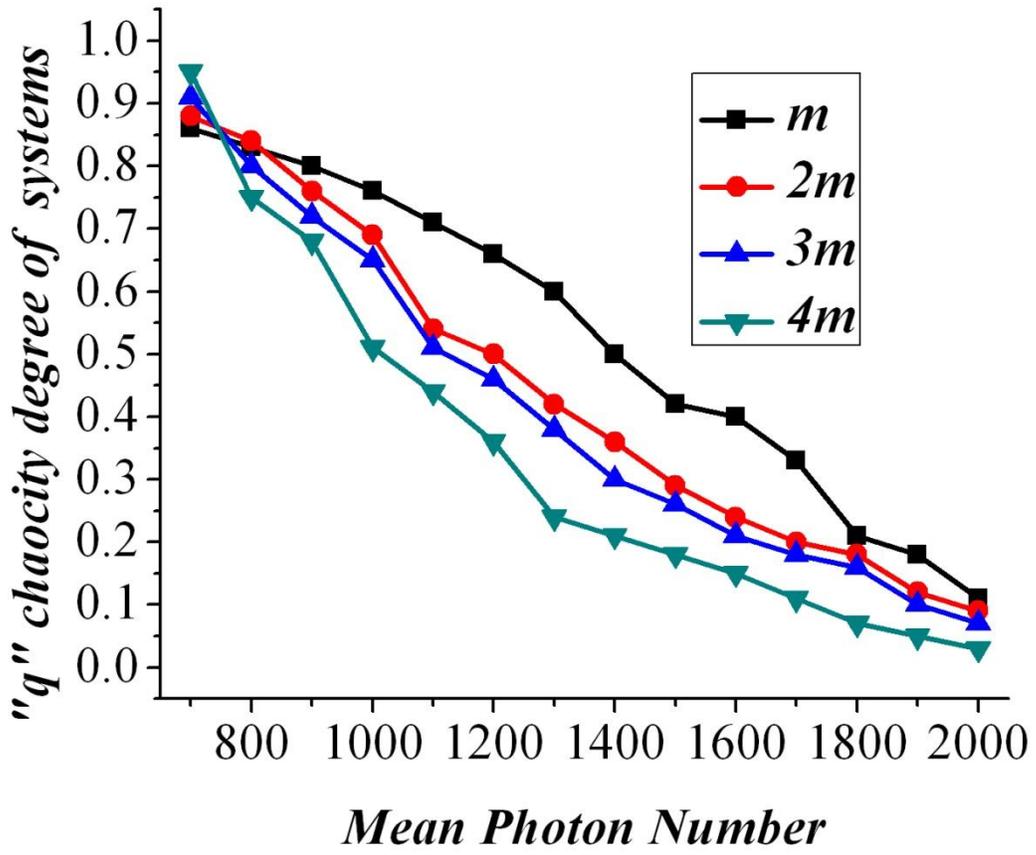